\documentclass[prd,amsmath,amssymb,preprint]{revtex4}
\usepackage{graphicx}
\usepackage{feynmf}
\usepackage{dcolumn}
\usepackage{bm}
\usepackage{longtable}

\begin{document}
\title{Relativistic calculations of the isotope shifts in 
highly charged Li-like ions}
\author{N.~A.~Zubova$^{1,2}$, Y.~S.~Kozhedub$^{1,2}$, V.~M.~Shabaev$^{1}$, I.~I.~Tupitsyn$^{1}$,  A.~V.~Volotka$^{1,3}$, G.~Plunien$^{3}$, C.~Brandau$^{4,5,6}$, and Th.~St\"ohlker$^{4,7,8}$}
\affiliation{$^1$Department of Physics, St. Petersburg State University, Ulianovskaya 1, Petrodvorets, St.~Petersburg
198504, Russia \\
$^2$SSC RF ITEP of NRC ``Kurchatov Institute'', Bolshaya Cheremushkinskaya 25, Moscow, 117218, Russia\\
$^3$Institut f\"ur Theoretische Physik, TU Dresden, Mommsenstrasse 13, Dresden, D-01062, Germany\\
$^4$ GSI Helmholtzzentrum f\"ur Schwerionenforschung GmbH, D-64291 
Darmstadt, Germany\\
$^5$  ExtreMe Matter Institute EMMI and Research Division, GSI 
Helmholtzzentrum f\"ur Schwerionenforschung, D-64291 Darmstadt, Germany\\
$^6$ Institut f\"ur Atom- und Molek\"ulphysik, Justus-Liebig-University Giessen, Leihgesterner Weg 217, D-35392 Giessen, Germany\\
$^7$ Helmholtz-Institut Jena, D-07743 Jena, Germany\\
$^8$ Institut f\"ur Optik und Quantenelektronik, 
Friedrich-Schiller-Universit\"at Jena, D-07743 Jena, Germany}
\date{\today}
\begin{abstract}
Relativistic calculations of the isotope shifts of energy levels in highly charged Li-like ions are performed.
The nuclear recoil (mass shift) contributions are calculated by merging the perturbative and large-scale configuration-interaction Dirac-Fock-Sturm 
(CI-DFS) methods. The nuclear size (field shift) contributions are evaluated by the CI-DFS method including the electron-correlation, Breit, 
and QED corrections.
The nuclear deformation and nuclear polarization corrections to the  isotope shifts in Li-like neodymium, thorium, and uranium 
are also considered.
The results of the calculations are compared with the theoretical values obtained with other methods.

\end{abstract}

\pacs{Valid PACS appear here}
\maketitle

\section{Introduction}
In the last years a great progress was achieved in experimental studies of the isotope shifts in highly charged ions \cite{Orts_2006,Brandau_2008}.
In Ref. \cite{Orts_2006} the isotope shift in B-like argon was measured employing laser spectroscopic methods at EBIT.
This experiment provided first tests of the relativistic theory of the nuclear recoil effect with highly charged ions \cite{Tupitsyn_2003}. 
In Ref. \cite{Brandau_2008} the measurements of the isotope shifts in dielectronic recombination spectra for Li-like neodymium ions with $A$=142 and $A$=150 
allowed determination of the nuclear charge radius difference. The accuracy of this experiment was also sensitive 
to the relativistic nuclear recoil contribution. Moreover, in Refs. {\cite{Brandau_2009,Brandau_2010,Brandau_2013}} it was demonstrated that the DR experiments at GSI can be extended to radioactive isotopes with a lifetime longer than about 10 s. It is expected that with the new FAIR facilities \cite{fair} in Darmstadt the isotope shift measurements 
in heavy ions will be improved in accuracy by an order of magnitude. From the theoretical side, to meet this accuracy one needs
to evaluate the nuclear size (field shift) and nuclear recoil (mass shift) contributions, including the relativistic and QED effects.

High-precision calculations of the mass shifts in highly charged Li-like ions were performed in Ref. \cite{Kozhedub_2010}, where the nuclear recoil contributions obtained 
within the Breit approximation (non-QED terms) were combined  with the related terms obtained using the relativistic theory beyond 
the Breit approximation (QED terms). The QED contributions were evaluated to zeroth order of the $1/Z$ perturbation theory
($Z$ is the nuclear charge number), while the Breit-approximation calculations were performed using the configuration-interaction
Dirac-Fock-Sturm (CI-DFS) method {\cite{Tupitsyn_2003}}. An independent calculation of the non-QED mass shifts was presented in Ref. {\cite{Li_2012}}. The results of this calculation, that was based on the multiconfiguration
Dirac-Fock (MCDF) method, agree with those from Ref. {\cite{Kozhedub_2010}} for low- and middle-Z ions. However, there is some discrepancy in the results for heavy Li-like ions.
Therefore, it would be very important to calculate the relativistic nuclear recoil contributions 
using a different approach. To this end, in the present paper we develop a method which merges the perturbative and CI-DFS calculations.
Namely, we calculate the nuclear recoil contributions within the Breit approximation to zeroth and first orders in $1/Z$
and add the related contributions of second and higher orders in $1/Z$, obtained using the CI-DFS method.
For checking purposes, we also perform the perturbative calculations starting with effective local potentials
that partly include the electron-electron interaction effects. Although the calculations of the $1/Z$ nuclear recoil contributions are restricted to the Breit approximation, the developed method can be straightforwardly extended beyond this approximation. The obtained non-QED results are combined with the corresponding QED contributions of the zeroth order in $1/Z$ to get the most accurate theoretical data for the mass shifts in highly charged Li-like ions.
In addition, the field shifts are calculated in the framework of the Dirac-Coulomb-Breit Hamiltonian. These calculations, being performed by the CI-DFS method,
are compared with the corresponding MCDF calculations of Ref. {\cite{Li_2012}}. The QED corrections to the field shifts are also evaluated. In addition, we consider the nuclear 
deformation and nuclear polarization corrections to the isotope shifts for Li-like neodymium, thorium, and uranium. As the result, the most precise theoretical values of the isotope shifts for the $2p_{1/2}-2s$ and $2p_{3/2}-2s$ transitions in Li-like ions are presented. 

The relativistic units ($\hbar=c=1$) are used throughout the paper. 

\section{Relativistic nuclear recoil effect}
Full relativistic theory of the nuclear recoil effect can be formulated only within quantum 
electrodynamics \cite{Shabaev_1985,Shabaev_1988,pac_95,Artemyev_1995, Shabaev_1998, Adkins_2007}. However, the lowest-order relativistic nuclear recoil corrections can be calculated within the Breit approximation employing the operator \cite{Shabaev_1985, Shabaev_1988, Palmer_1987}:
\begin{eqnarray}   
\label{rec1}
 H_M &=& \frac{1}{2M}\sum_{i,k}\Bigl[
{\vec{p_i}}\cdot {\vec{p_k}} -2\vec{D_i}
\cdot\vec{p_k}
\Bigr],
\end{eqnarray}
 where the indices $i$ and $k$ numerate the atomic electrons, Б┐≈$\vec{p}$ is the momentum operator, $\vec{\alpha}$ incorporates the Dirac
 matrices, and $\vec{D}$ is given by:
\begin{equation}
\vec{D}=\frac{\alpha Z}{2r}\Bigl[\vec{\alpha}+\frac{(\vec{\alpha}\cdot\vec{r})\vec{r}}{r^2}\Bigr].
\end{equation}
The nuclear recoil operator (\ref{rec1}) can be written as a sum:

\begin{equation}
H_M = H_{\rm{NMS}}+H_{\rm{RNMS}}+H_{\rm{SMS}}+H_{\rm{RSMS}},
\end{equation}
where 
\begin{equation}   \label{rec_NMS}
 H_{\rm{NMS}} = \frac{1}{2M}\sum_{i}\vec{p_i}^2\, 
\end{equation}
is normal mass shift (NMS) operator,
\begin{equation}   \label{rec_RMS}
H_{\rm{RNMS}} = -\frac{1}{2M}\sum_{i}\frac{\alpha Z}{r_i}\Bigl[\vec{\alpha_i}+\frac{(\vec{\alpha_i}\cdot\vec{r_i})\vec{r_i}}{r_i^2}\Bigr]\cdot\vec{p_i}\, 
 \end{equation}
is relativistic normal mass shift (RNMS) operator,
\begin{equation}   \label{rec_SMS}
 H_{\rm{SMS}} = \frac{1}{2M}\sum_{i\neq k}{\vec{p_i}\cdot \vec{p_k}}\, 
\end{equation}
is specific mass shift (SMS) operator, and
\begin{equation}   \label{rec_RSMS}
H_{\rm{RSMS}} = -\frac{1}{2M}\sum_{i\neq k}\frac{\alpha Z}{r_i}\Bigl[\vec{\alpha_i}+\frac{(\vec
{\alpha_i}\cdot\vec{r_i})\vec{r_i}}{r_i^2}\Bigr]\cdot\vec{p_k}\,
\end{equation}
is relativistic specific  mass shift (RSMS) operator.

Analytical calculations of the expectation values of the operators $H_{\rm{NMS}}$ and $H_{\rm{RNMS}}$ with the Dirac-Coulomb wave functions were performed in Ref. \cite{Shabaev_1985}. 
In Ref. \cite{Shabaev_1994}, the operator $H_{\rm{M}}$ was used to evaluate the lowest-order relativistic nuclear recoil corrections to energy levels of 
He- and Li-like ions to zeroth order in $1/Z$ (that corresponds to independent electron approximation).
Nowadays, this operator is widely used in relativistic calculations of the nuclear recoil
effect using the configuration-interaction and multiconfiguration Dirac-Fock methods 
\cite{Tupitsyn_2003, Kozhedub_2010, Li_2012, Kozlov_2007, Gaidamauskas_2011, 
Li_2013, Naze_2013}. It is known, however, that these methods
can have a rather poor convergence in calculations of the specific mass shift. Moreover, the CI-DF and MCDF methods can not be 
adopted to account for the QED nuclear recoil contribution,
which becomes very significant for heavy ions (see, e.g., Ref. \cite{Kozhedub_2010}). In the present paper 
we develop the perturbative approach to calculations of the interelectronic-interaction corrections to the mass shifts.
Although the perturbative calculations presented below are restricted to the Breit approximation, 
the developed approach has a potential to be extended to the full relativistic treatment.

To derive the nuclear recoil contributions to the binding energies of Li-like ions by perturbation theory, we use the two-time 
Green's function method \cite{Shabaev_2002} with the $(1s)^2$
shell regarded as belonging to a redefined vacuum. The energy shift of a level $a$ (valence state) due to all perturbative interactions is given by

\begin{equation}
\label{eq01}
\Delta E_a=\frac{\frac{1}{2\pi i} \displaystyle \oint \limits_{\Gamma}{dE \Bigl( E-E_a^{(0)} \Bigr) \Delta g_{aa}(E)}}{1+\frac{1}{2\pi i}
 \displaystyle \oint \limits_{\Gamma}{dE \Delta g_{aa}(E)}},
\end{equation}
where $\Delta g_{aa}(E)=g_{aa}(E)-g_{aa}^{(0)}(E)$, $g_{aa}(E)$ is the Fourier transform of the two-time Green function, 
projected on the unperturbed state $a$, $g_{aa}^{(0)}(E)={1}/{(E-E_a^{(0)})}$ is the unperturbed value of $g_{aa}(E)$, and $E_a^{(0)}$ is the unperturbed energy of the $a$ state, which in the case under consideration is simply equal to the Dirac energy of the valence electron: $E_a^{(0)}=\varepsilon_a$.
The contour $\Gamma$ surrounds the level $a$ and keeps outside all other singularities of $\Delta g_{aa}(E)$. It is oriented anticlockwise. The Green function $g_{aa}(E)$ is constructed by perturbation theory 
according to Feynman's rules given in Refs. \cite{Shabaev_1998,Shabaev_2002}. Since we restrict our calculation to the Breit approximation, we consider all the photon propagators in the Coulomb gauge at zero energy transfer ($\omega=0$) and restrict the summations over the intermediate electron states to the positive energy spectrum. In addition, we neglect the two-transverse photon nuclear recoil contributions \cite{Shabaev_1998}.

\begin{figure}[h]
\begin{minipage}[h]{0.49\linewidth}
\center{\includegraphics[width=0.5\linewidth]{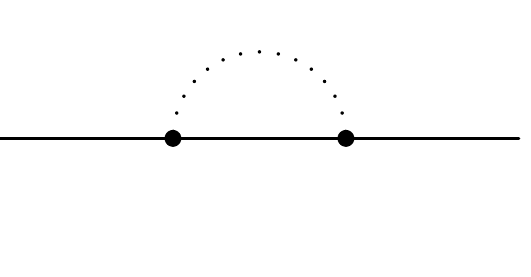}}
\end{minipage}
\vfill
\begin{center}
\begin{minipage}[h]{0.49\linewidth}
\center{\includegraphics[width=0.5\linewidth]{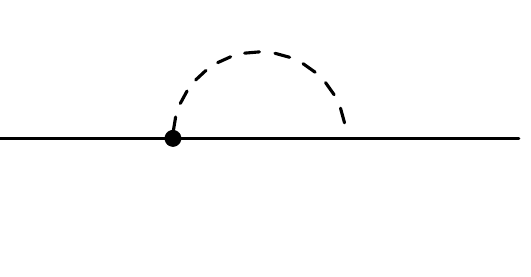} }
\end{minipage}
\hfill
\begin{minipage}[h]{0.49\linewidth}
\center{\includegraphics[width=0.5\linewidth]{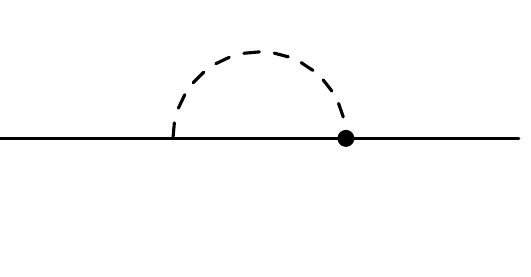}}
\end{minipage}
\end{center}
\caption{Feynman diagrams representing the lowest-order nuclear recoil corrections. The dotted line denotes the Coulomb recoil interaction that leads to the 
$\rm{NMS}$ and $\rm{SMS}$ contributions, the dashed line indicates the one-transverse-photon recoil interaction that leads to the $\rm{RNMS}$ and $\rm{RSMS}$ contributions.}
\label{fig1}
\end{figure}

To zeroth order in $1/Z$, the nuclear recoil corrections are defined by diagrams presented in Fig. \ref{fig1}. In these diagrams, in accordance with Refs. \cite{Shabaev_1998,Shabaev_2002}, the dotted line ended by bold wiles at both sides denotes the ``Coulomb recoil'' interaction that leads to the $\rm{NMS}$ and $\rm{SMS}$ contributions. The dashed line ended by a bold wile at one side designates the ``one-transverse-photon recoil'' interaction that leads to the $\rm{RNMS}$ and $\rm{RSMS}$ contributions. For the Coulomb recoil diagram (Fig. 1) one easily finds
\begin{equation}
\label{eq_4a}
\Delta g_{aa}^{(1)}=\frac{1}{{(E-\varepsilon_a)}^2}\frac{1}{M}\frac{i}{2\pi}
\int{d\omega \sum_n{\frac{{\langle a \mid \vec{p} \mid n \rangle}{\langle n \mid \vec{p} \mid a \rangle}}{E-\omega-\varepsilon_n+i{\eta}_{n}0}}},
\end{equation}
 where $\eta_n=\varepsilon_n-\varepsilon_{\rm{F}}$ and $\varepsilon_{\rm{F}}$ is the Fermi energy, which is chosen to be higher than the one-electron closed-shell energies and lower than the energies of the one-electron valence states. Using the identity
\begin{equation}
\label{eq_4a1}
\frac{1}{x \pm i0}={\mathcal P}\frac{1}{x} \mp {\pi{i}\delta(x)}  
\end{equation}
and formula (\ref{eq01}) to the first order, we get
\begin{equation}
\label{eq_4a2}
\Delta E_{\rm{Coul}}=\frac{1}{2M}\sum_{\varepsilon_n>\varepsilon_F}{{\mid \langle a \mid \vec{p} \mid n \rangle \mid}^2}-\frac{1}{2M}\sum_{\varepsilon_n<\varepsilon_F}{{\mid \langle a \mid \vec{p} \mid n 
\rangle \mid}^2}.
\end{equation}
This expression is conveniently divided into one and two-electron parts:
\begin{equation}
\label{eq_4a3}
\Delta E_{\rm{Coul}}=\Delta E^{\rm{(one-el)}}_{\rm{Coul}}+\Delta E^{\rm{(two-el)}}_{\rm{Coul}},
\end{equation}

\begin{equation}
\label{eq_4a4}
\Delta E_{\rm{Coul}}^{\rm{(one-el)}}=\frac{1}{2M}\sum_{\varepsilon_n>0}{{\mid \langle a \mid \vec{p} \mid n 
\rangle \mid}^2}-\frac{1}{2M}\sum_{\varepsilon_n<0}{{\mid \langle a \mid \vec{p} \mid n
\rangle \mid}^2},
\end{equation}

\begin{equation}
\label{eq_4a4}
\Delta E_{\rm{Coul}}^{\rm{(two-el)}}=-\frac{1}{M}\sum_{0<\varepsilon_c<\varepsilon_F}
{{\mid \langle a \mid \vec{p} \mid c
\rangle \mid}^2}.
\end{equation}
These formulas give the exact value of the Coulomb-recoil contribution within the full relativistic approach to zeroth order in $1/Z$. To separate the Breit-approximation term, we represent the one-electron contribution as follows

\begin{equation}
\label{eq_4a5}
\Delta E_{\rm{Coul}}^{\rm{(one-el)}}=\frac{1}{2M}
{\langle a \mid \vec{p}^2 \mid a
\rangle}-\frac{1}{M}\sum_{\varepsilon_n<0}{{\mid \langle a \mid \vec{p} \mid n
\rangle \mid}^2}.
\end{equation}
The first term in this equation gives the normal mass shift to zeroth order in $1/Z$, while the second term determines the QED part of the
one-electron Coulomb-recoil contribution. The expression $\Delta E_{\rm{Coul}}^{\rm{(two-el)}}$ defines the specific mass shift of zeroth order in $1/Z$. Therefore, we can write
\begin{equation}
\label{eq_4a6}
\Delta E_{\rm{NMS}}^{(0)}=\frac{1}{2M}
{\langle a \mid \vec p^2 \mid a\rangle},
\end{equation}
\begin{equation}
\label{eq_4a7}
\Delta E_{\rm{SMS}}^{(0)}=-\frac{1}{M}\sum_{0<\varepsilon_c<\varepsilon_F}
{{\mid \langle a \mid \vec{p} \mid c
\rangle \mid}^2},
\end{equation}
\begin{equation}
\label{eq_4a8}
\Delta E_{\rm{QED}}^{\rm{(0,Coul)}}=
-\frac{1}{M}\sum_{\varepsilon_n<0}{\mid \langle a \mid \vec{p} \mid n
\rangle \mid}^2,
\end{equation}
where the upper index $(0)$ corresponds to the zeroth order in $1/Z$. Performing similar calculations of the one-transverse-photon recoil contributions (Fig. 1 ) and keeping only the terms which correspond to the Breit approximation, we get
\begin{equation}
\label{eq_4a9}
\Delta  E_{\rm{RNMS}}^{(0)}=-\frac{1}{2M} {\langle a \mid (\vec{D} \cdot \vec{p}+\vec{p} \cdot \vec{D}) \mid a \rangle},
\end{equation}
\begin{equation}
\label{eq_4a10}
\Delta E_{\rm{RSMS}}^{(0)}=\frac{1}{M}\sum_{0<\varepsilon_c<\varepsilon_F}{({\langle a \mid \vec{p} \mid c
\rangle}{\langle c \mid \vec{D} \mid a
\rangle}+{\langle a \mid \vec{D} \mid c
\rangle}{\langle c \mid \vec{p} \mid a
\rangle})}.
\end{equation}

\noindent

Let us consider the electron-electron interaction corrections to the nuclear recoil effect. To first order in $1/Z$, the interelectronic-interaction corrections to the $\rm{NMS}$ and $\rm{SMS}$ contributions are defined by Feynman's diagrams presented in Fig. 2. In these diagrams, the wavy line indicates the electron-electron interaction taken in the Breit approximation:                                      
\begin{eqnarray}
\label{V12}                                                                    
V(1,2)=V_{C}(1,2)+V_{B}(1,2)
=\frac{\alpha}{r_{12}}
-\alpha\Bigr[
\frac{\vec \alpha_1\cdot{\vec \alpha_2 }}
{r_{12}}+\frac{1}{2}({\vec \nabla_1} \cdot {\vec \alpha_1})({\vec \nabla_2} \cdot {\vec \alpha_2})r_{12}\Bigr].  
\end{eqnarray}

\begin{figure}[h]
  \begin{minipage}[h]{0.49\linewidth}
\center{\includegraphics[width=0.5\linewidth]{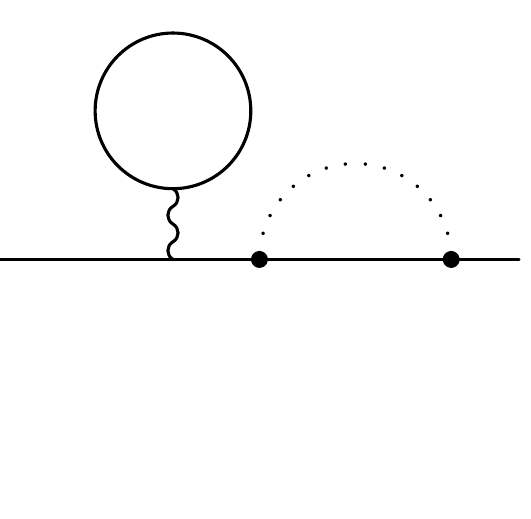} \\[-16mm] a}
\end{minipage}
\hfill
\begin{minipage}[h]{0.49\linewidth}
\center{\includegraphics[width=0.5\linewidth]{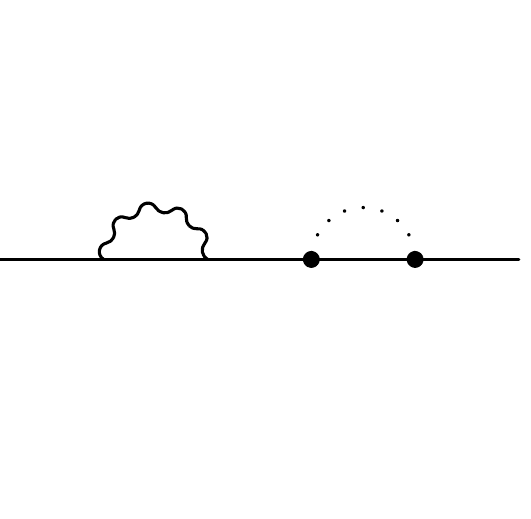} \\[-16mm] b}
\end{minipage}
\\[7mm]
\begin{minipage}[h]{0.49\linewidth}
\center{\includegraphics[width=0.5\linewidth]{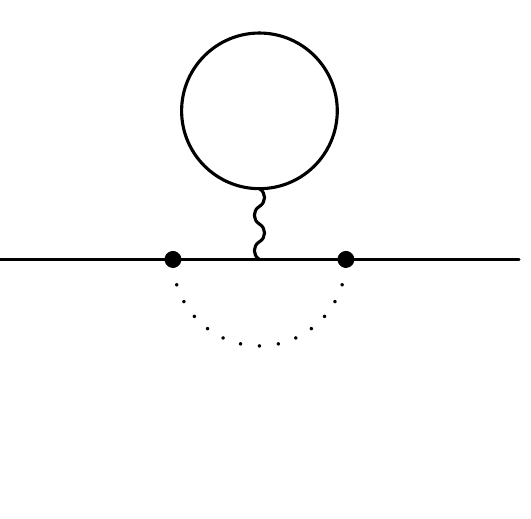} \\[-16mm] c}
\end{minipage}
\begin{minipage}[h]{0.49\linewidth}
\center{\includegraphics[width=0.5\linewidth]{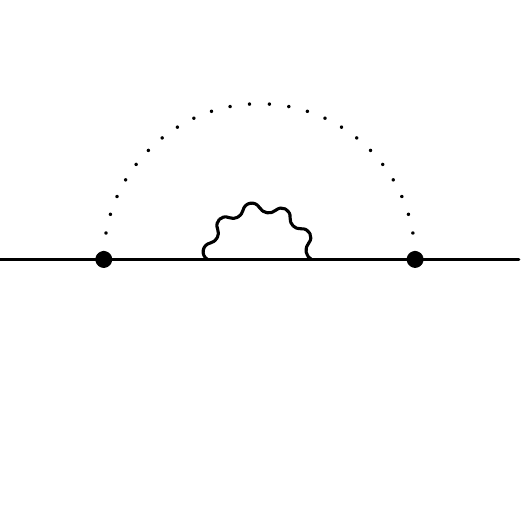} \\[-16mm] d}
\end{minipage}
\\[7mm]
\begin{minipage}[h]{0.49\linewidth}
\center{\includegraphics[width=0.5\linewidth]{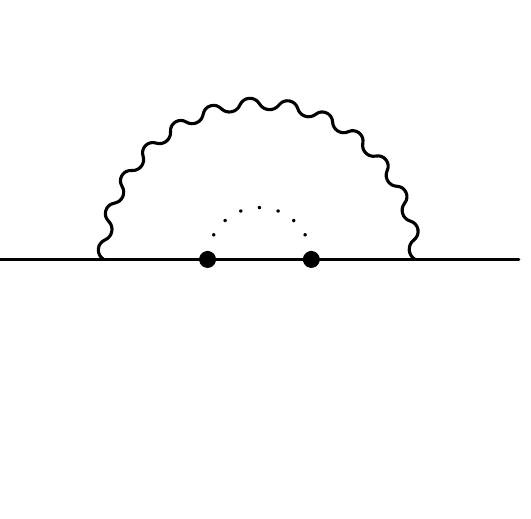} \\[-16mm] e}
\end{minipage}
\begin{minipage}[h]{0.49\linewidth}
\center{\includegraphics[width=0.5\linewidth]{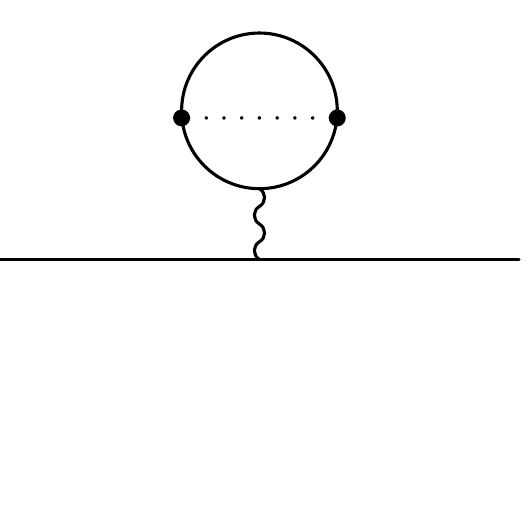} \\[-16mm] f}
\end{minipage}
\\[7mm]
\begin{minipage}[h]{0.49\linewidth}
\center{\includegraphics[width=0.5\linewidth]{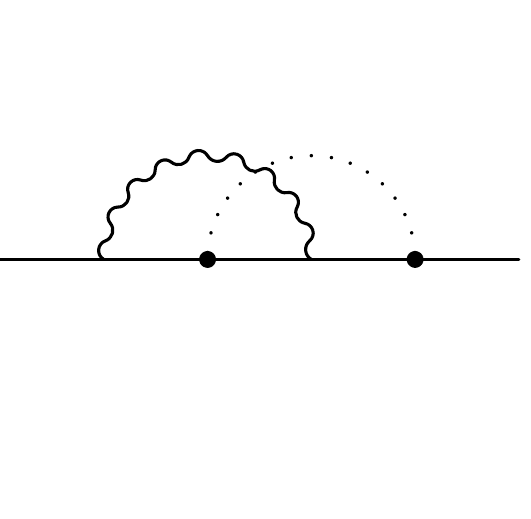} \\[-16mm] g}
\end{minipage}
\begin{minipage}[h]{0.49\linewidth}
\center{\includegraphics[width=0.5\linewidth]{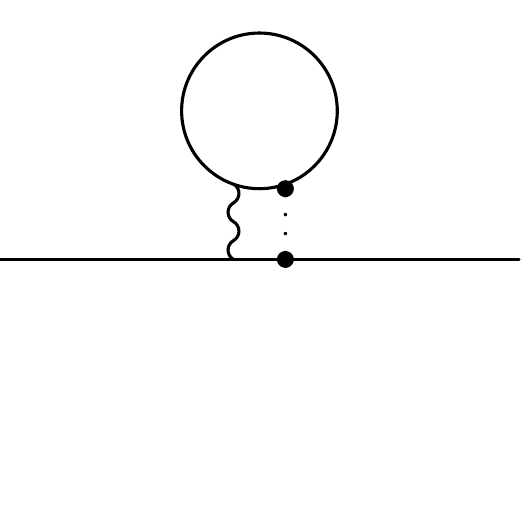} \\[-16mm] h}
\end{minipage}
\caption{Feynman diagrams representing the interelectronic-interaction corrections of the first order in 1/Z to the NMS or SMS contributions.}
\label{ris:image1}
\end{figure}
\noindent
In Fig. 2 we retain only those diagrams, which give nonzero contributions for a Li-like ion with one valence 
electron over the closed $(1s)^2$ shell and keep in mind the symmetric partners of the diagrams $a$, $b$, $g$, and $h$, which double the corresponding contributions. Similar diagrams, with the dotted line replaced by the dashed lines as in Fig. 1, determine the interelectronic-interaction corrections to the $\rm{RNMS}$ and $\rm{RSMS}$ contributions. The calculation of the diagram $a$ and the related partners using formula (\ref{eq01}) leads to the following expressions:  
\begin{equation}
\label{NMS_a}
\Delta E_{\rm{NMS}}^{(1,a)}=\frac{1}{M} \sum \limits_{\varepsilon_n>0}^{(\varepsilon_n \neq \varepsilon_a)}
\sum \limits_{0<\varepsilon_c<\varepsilon_F}\frac {1}{{\varepsilon}_a-{\varepsilon}_{n}}{\langle a c \mid V\mid nc \rangle}{\langle n \mid \vec{p}^2 \mid a\rangle}, 
\end{equation}

\begin{equation}
\label{RNMS_a}
  \Delta E_{\rm{RNMS}}^{(1,a)}=-\frac{1}{M} \sum \limits_{\varepsilon_n >0}^{(\varepsilon_n \neq \varepsilon_a)} \sum \limits_{0<\varepsilon_c<\varepsilon_F}
 \frac{1}{{\varepsilon}_a-{\varepsilon}_{n}} {\langle a c \mid V\mid nc\
 \rangle}{\langle n \mid (\vec{D}\cdot \vec{p}+\vec{p}\cdot \vec{D}) \mid
a\rangle} 
\end{equation}

\begin{equation}
\label{SMS_a}
\Delta E_{\rm{SMS}}^{(1,a)}=
-\frac{2}{M} \sum \limits_{\varepsilon_n >0}^{(\varepsilon_n \neq \varepsilon_a)} \sum \limits_{0<\varepsilon_c<\varepsilon_F} 
\sum \limits_{0<\varepsilon_c'<\varepsilon_F}
\frac {1}{{\varepsilon}_a-{\varepsilon}_{n}} {\langle a c \mid V\mid nc \rangle}{\langle n \mid \vec{p} \mid
c'\rangle}{\langle c' \mid \vec{p} \mid a \rangle} ,
\end{equation}

\begin{eqnarray}
\label{RSMS_a}
\Delta E_{\rm{RSMS}}^{(1,a)}=
\frac{2}{M} \sum \limits_{\varepsilon_n >0}^{(\varepsilon_n \neq 
\varepsilon_a)} \sum \limits_{0<\varepsilon_c<        
\varepsilon_F} \sum \limits_{0<\varepsilon_c'<\varepsilon_F}
\frac {1}{{\varepsilon}_a-{\varepsilon}_{n}} 
{\langle a c \mid V\mid nc \rangle}\\
\notag
\times {\Bigl( {\langle n \mid \vec{p} \mid
c'\rangle}{\langle c' \mid \vec{D} \mid a \rangle}+{\langle n \mid \vec{D} \mid
c'\rangle}{\langle c' \mid \vec{p} \mid a \rangle}\Bigr)},
\end{eqnarray}
where the scalar products of the vectors are implicit. For the other diagrams ($b$-$h$), we give the explicit expressions for the NMS and SMS contributions only:

\begin{equation}
\label{NMS_b}
\Delta E_{\rm{NMS}}^{(1,b)}=-\frac{1}{M} \sum \limits_{\varepsilon_n >0}^{(\varepsilon_n 
\neq \varepsilon_a)}
\sum \limits_{0<\varepsilon_c<\varepsilon_F}
\frac {1}{{\varepsilon}_a-{\varepsilon}_{n}} 
{\langle a c \mid V \mid c n \rangle} {\langle n \mid \vec{p}^2 \mid a \rangle},
\end{equation}

\begin{equation}
\label{SMS_b}
\Delta E_{\rm{SMS}}^{(1,b)}=\frac{2}{M}\sum \limits_{ \varepsilon_n >0}^{(\varepsilon_n \neq \varepsilon_a)} \sum \limits_
{0<\varepsilon_c<\varepsilon_F} 
\sum \limits_{0<\varepsilon_c'<\varepsilon_F}{\frac {1}{{\varepsilon}_a-{\varepsilon}_{n}}}
{\langle a c \mid V\mid c n \rangle}
{\langle n \mid \vec{p} \mid c'\rangle}{\langle c' \mid \vec{p} \mid a \rangle},
\end{equation}

\begin{equation}
\label{NMS_c}
\Delta E_{\rm{NMS}}^{(1,c)}=0,
\end{equation}

\begin{eqnarray}
\label{SMS_c}
\Delta E_{\rm{SMS}}^{(1,c)}
=\frac{1}{M}\sum \limits_{\varepsilon_n>\varepsilon_F}\sum \limits_
{0<\varepsilon_c<\varepsilon_F} \sum \limits_{0<\varepsilon_c'<\varepsilon_F} 
{\frac{1}{{\varepsilon}_n-{\varepsilon}_c}
\Bigl({\langle a \mid \vec{p} \mid n \rangle}
{\langle c \mid \vec{p} \mid a \rangle}
{\langle n c' \mid V\mid c c'\rangle}}\\
\notag
+
{{\langle a \mid \vec{p} \mid c \rangle}
{\langle n \mid \vec{p} \mid a \rangle}
{\langle c c' \mid V\mid n c'\rangle}\Bigr)},
\end{eqnarray}

\begin{equation}
\label{NMS_d}
\Delta E_{\rm{NMS}}^{(1,d)}=0,
\end{equation}

\begin{eqnarray}
\label{SMS_d}
\Delta E_{\rm{SMS}}^{(1,d)}
=-\frac{1}{M}  \sum \limits_{\varepsilon_n>\varepsilon_F}\sum \limits_{0<\varepsilon_c<\varepsilon_F} \sum \limits_{0<\varepsilon_c'<\varepsilon_F}
{\frac{1}{\varepsilon_n-\varepsilon_c}
\Bigl(
{\langle a \mid \vec{p} \mid n \rangle} 
{\langle c \mid \vec{p} \mid a \rangle}{\langle n c' \mid V\mid c' c\rangle}}\\
\notag
+
{{\langle a \mid \vec{p} \mid c \rangle} {\langle n \mid \vec{p} \mid a \rangle}
{\langle c c' \mid V\mid c' n\rangle} \Bigr)}, 
\end{eqnarray}

\begin{eqnarray}
\label{NMS_e}
\Delta E_{\rm{NMS}}^{(1,e)}= \frac{1}{M} \sum \limits_{\varepsilon_n>\varepsilon_F}
\sum \limits_{0<\varepsilon_c<\varepsilon_F}{\frac{1}{\varepsilon_n-\varepsilon_c}{\langle ac \mid V\mid na\rangle}
{\langle n \mid \vec{p}^2 \mid c \rangle}},
\end{eqnarray}

\begin{equation}
\label{SMS_e}
\Delta E_{\rm{SMS}}^{(1,e)}=0,
\end{equation}

\begin{equation}
\label{NMS_f}
\Delta E_{\rm{NMS}}^{(1,f)}= \frac{1}{M}\sum \limits_{\varepsilon_n>\varepsilon_F}
\sum \limits_{0<\varepsilon_c<\varepsilon_F} {{\langle a c \mid V\mid a n \rangle}\frac{1}{\varepsilon_c-\varepsilon_n}{\langle n \mid \vec{p}^2 \mid c \rangle}},
\end{equation}

\begin{equation}
\label{SMS_f}
\Delta E_{\rm{SMS}}^{(1,f)}=0,
\end{equation}

\begin{equation}
\label{NMS_g}
\Delta E_{\rm{NMS}}^{(1,g)}=0,
\end{equation}

\begin{eqnarray}
\label{SMS_g}
\Delta E_{\rm{SMS}}^{(1,g)}=-\frac{2}{M} \Bigl(
\sum \limits_{0<\varepsilon_c<\varepsilon_F}\sum \limits_{0<\varepsilon_c'<\varepsilon_F}
\sum \limits_{\varepsilon_n>\varepsilon_F}{\frac{{\langle a \mid \vec{p} \mid c' \rangle}{\langle n \mid \vec{p} \mid c \rangle}}
{\varepsilon_a+\varepsilon_n-\varepsilon_c-\varepsilon_c'}{\langle c'c\mid V\mid na\rangle}}\\
\notag
+\sum \limits_{0<\varepsilon_c<\varepsilon_F}\sum \limits_{\varepsilon_n>0}
\sum \limits_{\varepsilon_n'>\varepsilon_F}^{(\varepsilon_n+\varepsilon_n' \neq \varepsilon_a+\varepsilon_c)}
{\frac{{\langle a \mid \vec{p} \mid n' \rangle}{\langle c \mid \vec{p} \mid n \rangle}}{\varepsilon_a+\varepsilon_c-
\varepsilon_n-\varepsilon_n'}{\langle n'n\mid V\mid ca\rangle}}
\Bigr),
\end{eqnarray}

\begin{equation}
\label{NMS_h}
\Delta E_{\rm{NMS}}^{(1,h)}=0,
\end{equation}

\begin{eqnarray}
\label{SMS_h}
\Delta E_{\rm{SMS}}^{(1,h)}=\frac{2}{M} \Bigl(
\sum \limits_{\varepsilon_n>\varepsilon_F}\sum \limits_{\varepsilon_n'>\varepsilon_F}^{(\varepsilon_n+\varepsilon_n' \neq \varepsilon_a+\varepsilon_c)}\sum 
\limits_{0<\varepsilon_c<\varepsilon_F}{\frac{{\langle a \mid \vec{p} \mid n' \rangle}{\langle c \mid 
\vec{p} \mid n \rangle}{\langle nn'\mid V \mid ca\rangle}}
{\varepsilon_a+\varepsilon_c-\varepsilon_n-\varepsilon_n'}}\\
\notag
+\sum \limits_{0<\varepsilon_c<\varepsilon_F} \sum \limits_{0<\varepsilon_c'<\varepsilon_F}
\sum \limits_{\varepsilon_n>0}{\frac{{\langle a \mid \vec{p} \mid c' \rangle}{\langle n \mid \vec{p} \mid c 
\rangle}}{\varepsilon_a+\varepsilon_n-
\varepsilon_c-\varepsilon_c'}{\langle cc'\mid V\mid na\rangle}}
\Bigr).
\end{eqnarray}
The RNMS and RSMS contributions corresponding to the diagrams $b$-$h$ are easily obtained from Eqs. (\ref{NMS_b})-(\ref{SMS_h}) by replacing $\vec{p}$ with $\vec{p}-\vec{D}$ and keeping only the terms which contain the $\vec{p}$ and $\vec{D}$ operators.
After the angular integration with help of the Eckart-Wigner theorem, the numerical calculations of the expressions (\ref{NMS_a})-(\ref{SMS_h}) are performed employing the dual-kinetic-balance (DKB) finite basis set method \cite{Shabaev_2004} with the basis functions
constructed from B-splines \cite{Johnson_1988}.
The calculations have been carried out for extended nuclei. The Fermi model was used to describe the nuclear charge 
distribution and the nuclear charge radii were taken from Refs. \cite{Angeli_2013,Kozhedub_2008}. 

To evaluate the nuclear recoil corrections of the second and higher orders in $1/Z$, we used the CI-DFS method. With this method we calculated the total nuclear recoil contributions within the Breit approximation, including the Coulomb and Breit electron-electron interaction projected on the positive energy states. This was done by evaluating the expectation value of the nuclear recoil operator (\ref{rec1}) with the CI-DFS wave function. To separate terms of different orders in $1/Z$, the electron-electron interaction operator was taken in a form:
\begin{equation}
V(\lambda)=\lambda V,
\end{equation}
where $V$ is given by equation (\ref{V12}) and $\lambda$ is a scaling parameter. For small $\lambda$, the nuclear recoil contribution can be expanded in powers of $\lambda$:
\begin{equation}
E(\lambda)=E_0+E_1 \lambda +\sum \limits_{k=2}^{\infty}{E_k \lambda^k},
\end{equation}   
where
\begin{equation}
\label{Ek}
E_k=\frac{1}{k!}\frac{d^k}{d\lambda^k}E(\lambda)_{|{\lambda=0}}.
\end{equation}
The second- and higher-order contribution $E_{\ge2}=\sum \limits_{k=2}^{\infty}{E_k}$ is calculated as
\begin{equation}
E_{\ge 2}=E(\lambda=1)-E_0-E_1,
\end{equation} 
where the terms $E_0$ and $E_1$ are determined numerically according to equation (\ref{Ek}).

Finally, one should consider the nuclear recoil contributions beyond the Breit approximation ($\rm{QED}$ nuclear recoil terms). The $\rm{QED}$ calculations of the nuclear recoil effect for highly charged ions to zeroth order in $1/Z$ were performed in Refs. {\cite{Artemyev_1995,Artemyev2_1995,Adkins_2007}} for point charged nuclei and in Refs. {\cite{Shabaev_1998_2,Shabaev_1999,Kozhedub_2010}} for extended nuclei. In the present paper, to get the QED nuclear recoil corrections, we interpolated the corresponding data from Ref. {\cite{Kozhedub_2010}}.

\section{Finite nuclear size effect} 
The finite size of atomic nuclei leads to the field shifts of the energy levels. The nuclear charge distribution is usually approximated by the spherically-symmetric Fermi model:
\begin{equation}
\label{rho}
\rho(r,R)=\frac{N}{1+{\rm{exp}}[(r-c)/a]},
\end{equation}
where the parameter $a$ is generally fixed to be $a=2.3/(4{\rm ln}3)$ fm and the parameters $N$ and $c$ are determined using the given value of the root-mean-square ($\rm{rms}$) nuclear charge radius $R=\langle r^2 \rangle^{1/2}$ and the normalization condition: $\int{d\vec{r} \rho({r},R)}=1$. 
The potential induced by the nuclear charge distribution $\rho(r,R)$ is defined as 
\begin{equation}
\label{Vn}
 V_{N}(r,R)= -4\pi \alpha Z \int\limits_{0}^{\infty} {dr' r^2 \rho (r',R) \frac{1}{r_{>}}},
\end{equation}
where $r_{>}=\rm{max}(r,r')$. 
The isotope field shift within the Breit approximation can be obtained by solving the Dirac-Coulomb-Breit equation with the potential (\ref{Vn}) for two different isotopes and taking the corresponding energy difference.

Since the finite nuclear size effect is mainly determined by the $\rm{rms}$ nuclear charge radius (see, e.g., Ref. \cite{Shabaev_1993}), the energy difference between two isotopes can be approximated as
\begin{equation}
\label{FS_0}
\delta E_{FS} = {F}\delta \langle r^2 \rangle,
\end{equation}
where $F$ is the field shift factor and $\delta \langle r^2 \rangle$ is the 
mean-square charge radius difference. 
In accordance with this definition, the $F$-factor can be calculated by
\begin{equation}
\label{FS_1}
F=
\frac{dE(R)}{d\langle r^2 \rangle} 
\end{equation}
or, using the Hellmann-Feynman theorem, by
\begin{equation}
\label{FS_2}
F= \langle \psi \mid \sum_{i} \frac{dV_{N}(r_{i},R)}{d\langle r^2 \rangle} \mid  \psi \rangle,
\end{equation}
where $\psi$ is the wave function of the state under consideration and the index $i$ runs over all atomic electrons. If we neglect the variation of the electronic density inside the nucleus, we get (see, e.g., Refs. \cite{Torbohm_1985,Li_2012,Naze_2013}):
\begin{equation}
\label{FS_3}
F=\frac{2 \pi}{3} \alpha Z {\mid \psi(0) \mid}^2.
\end{equation}
In what follows, the values of $F$ calculated by formulas (\ref{FS_1}),(\ref{FS_2}), and (\ref{FS_3}) will be referred as obtained by methods 1, 2, and 3, respectively.
In addition to the $\rm{FS}$ evaluated with the DCB Hamiltonian, one should account for the $\rm{QED}$ corrections to the field shift. Approximately, these corrections can be evaluated using analytical formulas from Ref. \cite{Milstein_2004}.
The results obtained by these formulas for $s$ and $p_{1/2}$ states are in a fair agreement with the accurate numerical calculations performed for H-like ions in Ref. {\cite{Yerokhin_2011}}.

\section{Results and discussions}
The nuclear recoil contributions are conveniently expressed in terms of the K-factor defined by 
\begin{equation}
\Delta E=\frac{K}{M}.
\end{equation}
It follows that the isotope mass shift is given by
\begin{equation}
\label{MS_1}
\delta E_{MS}=\frac{K}{M_1}-\frac{K}{M_2}=-\frac{\delta M}{M_1 M_2}K,
\end{equation}
where $\delta M=M_1-M_2$ is the nuclear mass difference. In Table \ref{tab1} we present the contributions of individual diagrams to $K_{\rm{NMS}}$, $K_{\rm{SMS}}$, $K_{\rm{RNMS}}$, and $K_{\rm{RSMS}}$ for the $2p_j-2s$ transitions in Li-like uranium. In contrast to our previous paper \cite{Zubova_2013}, where the $\rm{NMS}$ and $\rm{RNMS}$ contributions were evaluated by summing over all intermediate electronic states, here we restrict all the summations to the positive-energy states only. The difference due to the negative-energy states contributes on the level of the $\rm{QED}$ recoil corrections of the first order in $1/Z$, which are beyond the scope of the present paper.

The values of $K_{\rm{{NMS}}}$, $K_{\rm{{RNMS}}}$, $K_{\rm{{SMS}}}$ , $K_{\rm{RSMS}}$ and $K_{\rm{QED}}$ for the $2p_{j}-2s$ transition energies in Li-like ions 
are presented in Tables \ref{tab2} and \ref{tab3} for $j=1/2$ and $j=3/2$, respectively. The nuclear recoil contributions of the zeroth and first orders in $1/Z$ are calculated by the perturbative approach while the corresponding terms of the second and higher orders in $1/Z$ are evaluated using the CI-DFS method. The total MS within the Breit approximation is given by the sum of the $\rm{NMS}$, $\rm{RNMS}$, $\rm{SMS}$, and $\rm{RSMS}$ contributions evaluated to all orders in $1/Z$. The obtained results are compared with the related theoretical data from Refs. {\cite{Kozhedub_2010,Li_2012}}.
The total MS values including the $\rm{QED}$ recoil contributions are also presented.
From Tables \ref{tab2} and \ref{tab3} it can be seen
that the present results obtained using the perturbative approach are in perfect agreement with the calculations based on the CI-DFS method \cite{Kozhedub_2010}. As to comparison with the MCDF calculations of Ref. {\cite{Li_2012}}, there exists some discrepancy for heavy ions. We note that this discrepancy is larger than the contribution of the second and higher orders in $1/Z$. 

For checking purposes, we have also performed the perturbative calculations starting with an effective potential, which includes both the Coulomb nuclear potential and the screening potential that partly accounts for the electron-electron interaction. The calculations are performed to the zeroth and first orders in $1/Z$ with four different potentials: Dirac-Slater ($\rm{DS}$) \cite{Slater_1951}, Kohn-Sham ($\rm{KS}$) \cite{Kohn_1965}, Perdew-Zunger ($\rm{PZ}$) \cite{Perdew_1981}, and local Dirac-Fock ($\rm{LDF}$) \cite{Shabaev_2005}.
All these potentials
 have been successfully used in calculations of highly charged ions (see, e.g., Refs.\cite{Sapirstein_2001, Artemyev_2007,Yerokhin_2007,Volotka_2008, Sapirstein_2011,Artemyev_2013} and references therein). To avoid the double counting, the interaction with the screening potential conterterm is accounted for perturbatively. The results of these calculations for Li-like uranium are presented in Table {\ref{tab4}}. For comparison, the perturbative results based on the Coulomb nuclear potential and the CI-DFS results of Ref. \cite{Kozhedub_2010} are given as well. It can be seen that the $\rm{DS}$, $\rm{KS}$, $\rm{PZ}$, and $\rm{LDF}$ perturbative results, which include the two lowest-order contributions only, are in a good agreement with the all-order Coulomb perturbative and CI-DFS calculations.
The especially good convergence is observed for the $\rm{LDF}$ potential. Thus, the CI-DFS results of Ref. \cite{Kozhedub_2010} are confirmed by fully independent perturbative calculations. 

In Table \ref{tab5} we present the total values of the mass shifts in the range $Z$=4 - 92. For $Z \ge  20$ the Breit-approximation values are obtained by merging the perturbative approach with the CI-DFS method, while for $Z<20$ the pure CI-DFS calculations are used. The QED contributions, being calculated in the independent-electron approximation (to zeroth order in $1/Z$) \cite{Kozhedub_2010}, become dominant for heavy ions. The uncertainty was evaluated as a quadratic sum of the uncertainty due to the CI-DFS calculation and the uncertainty due to uncalculated QED contributions of the first order in $1/Z$. The latter one was estimated as the QED contribution of zeroth order in $1/Z$ multiplied with a factor $2/Z$ (compared to Ref. {\cite{Kozhedub_2010}}, we use here a more conservative estimate).
As to the uncertainty of the CI-DFS calculation, it was determined by studying the convergence of the obtained results with respect to the number of the basis functions.
For further improvement of the theoretical accuracy in high-Z region, calculations of the QED corrections of the first order in $1/Z$ are needed. The perturbative approach developed in this paper can be considered as the first necessary step for such calculations.

To evaluate the $\rm{FS}$ constant with the DCB Hamiltonian we used the CI-DFS method. In Table \ref {tab6} we compare the non-QED $F$-factor, obtained by equations (\ref{FS_1}), (\ref{FS_2}) and (\ref{FS_3}) (methods 1, 2, and 3, respectively), for Li-like titanium ($Z=22$), 
neodymium ($Z=60$), and thorium ($Z=90$). 
The results of Ref. \cite{Li_2012}, where the method 3 was employed, are also 
presented. It can be seen that the last method leads to a rather poor accuracy for heavy ions. 
In case of Li-like thorium the discrepancy between the most precise 
result obtained by methods 1 and 2 and that obtained by method 3 amounts to about 10 \%. The discrepancy is much larger than the uncertainty due to 
neglecting the non-linear corrections to formula (\ref{FS_0}). This is confirmed by the data presented in Table \ref{tab7}. In this table
 the non-QED FS contributions to the isotope shift obtained by the direct calculation: $\delta E=E(R_1)-E(R_2)$, 
where $R_1$ and $R_2$ are the nuclear charge radii of the isotopes taken from Ref. {\cite{Angeli_2013}},
 are compared with the corresponding results calculated using the $F$-factor. 

Table \ref{tab8} presents the Dirac-Fock, Breit, electron-correlation, and QED contributions to the field-shift constant ${F}$ for the $2p_{1/2}-2s$ and $2p_{3/2}-2s$ transitions in Li-like titanium, neodymium, and thorium. The QED corrections to the nuclear size effect are evaluated using the approximate analytical formulas for H-like ions from Refs. \cite{Milstein_2004, Yerokhin_2011}.
This was done by multiplying the $s$-state QED correction factor $\Delta_s$ \cite{Milstein_2004, Yerokhin_2011} with the nuclear size effect on the total three-electron binding energy. The uncertainty of this evaluation was determined by comparing the obtained results for the vacuum-polarization correction with the related direct calculation and assuming the relative uncertainty of the total QED correction to be by 50 \% larger. 
These calculations demonstrate rather large values of the QED contributions to the field shift for heavy ions. In Tables \ref{tab9}  and  \ref{tab10} we present our total values of the $FS$ constant for the $2p_{1/2}-2s$ and $2p_{3/2}-2s$ transitions in Li-like ions in the range $Z$=4-92. The uncertainty was evaluated as a quadratic sum of the uncertainty due to a variation of the nuclear charge radius value taken from Ref. {\cite{Angeli_2013}}, the uncertainty due to the determination of the QED contributions discussed above, and the uncertainty due to a variation of the nuclear charge distribution which was estimated as the difference between the results obtained for the Fermi and homogeneously charged sphere models. 

Table \ref{tab11} presents individual contributions to the isotope shifts of the $2p_{1/2}-2s$ and $2p_{3/2}-2s$ transitions in Li-like $^{A}\rm{Nd}^{57+}$ ions with $A$=142 and $A$=150, which were measured in Ref. \cite{Brandau_2008}. In addition to the mass and field shifts, one has to account for the nuclear polarization \cite{Plunien_1991, Plunien_1991_corr, Nefiodov_1996,Volotka_press} and nuclear deformation \cite{Kozhedub_2008} effects. To evaluate the nuclear polarization effect, we used the approach \cite{Plunien_1991, Plunien_1991_corr} in which the many-body theory for
virtual nuclear excitations was incorporated with the bound-state QED for the
atomic electrons. For low-lying rotational and vibrational levels the nuclear
excitation energies and transition probabilities for $^{142}\rm{Nd}$ and $^{150}\rm{Nd}$ 
nuclei
were taken from Refs. \cite{Tuli_2000, Basu_2013}, respectively.
The contributions from the nuclear giant
resonances were evaluated utilizing phenomenological energy-weighted sum rules.
 To calculate the nuclear deformation correction, the standard spherically-symmetric Fermi model of the nuclear charge distribution (\ref{rho}) must be replaced by {\cite{Kozhedub_2008}}:
\begin{equation}
\rho({r})=\frac{1}{4\pi}\int{d\vec{n} \rho(\vec{r})},
\end{equation}
where $\rho(\vec{r})$ is the deformed Fermi distribution:
\begin{equation}
\rho(\vec{r})=\frac{N}{1+{\rm{exp}}\{[r-r_0(1+\beta_{20}Y_{20}({\Theta}))]/a \}},
\end{equation}
$Y_{20}({\Theta})$ is the spherical function and $\beta_{20}$ is the quadrupole deformation parameter. In accordance with Ref. \cite{Fricke_1995} we take $\beta_{20}=$0 for $A$=142 and $\beta_{20}=$0.28(5) for $A$=150, that leads to a non-zero  nuclear deformation effect for the $^{150}{\rm{{Nd}}}$ isotope only. The nuclear deformation correction is given by the difference between the nuclear size contributions evaluated with non-zero and zero values of $\beta_{20}$ for the same nuclear parameters ${\langle r^2 \rangle}^{1/2}$ and the atomic mass numbers. As it can be seen from Table {\ref{tab11}}, the nuclear polarization and deformation contributions to the isotope shifts are comparable with the QED corrections. The perfect agreement of the theoretical value of the isotope shift for the $2p_{1/2}-2s$ transition with the experiment \cite{Brandau_2008} should not be surprising since the 
mean-square charge radius difference  $\delta {\langle r^2 \rangle}=$1.36(1)(3) $\rm{fm^2}$ was determined from this comparison \cite{Brandau_2008}.

In Table \ref{tab12} we present the isotope shifts of the $2p_j-2s$ transitions in Li-like thorium with atomic numbers $A$=232 and $A$=230, and in Li-like uranium for two pairs of  even-even isotopes, 
 $^{238}{\rm{{U}}^{89+}}-^{236}{\rm{{U}}^{89+}}$ and $^{238}{\rm{{U}}^{89+}}-^{234}{\rm{{U}}^{89+}}$. The values of $\delta {\langle r^2 \rangle}^{1/2}$ are taken from Ref. {\cite{Angeli_2013}}. The mass and field shifts, including the QED corrections, are calculated as in the neodymium case. The nuclear polarization effect for thorium and uranium was evaluated in Refs. \cite{Plunien_1991, Plunien_1991_corr}. The nuclear deformation effect was calculated as in Ref. \cite{Kozhedub_2008}, using
the experimental \cite{Bemis_1973, Zumbro_1984} and theoretical \cite{Moeller_1995} data for the nuclear deformation parameters. The total uncertainty is mainly determined by the uncertainties of the nuclear deformation and polarization effects.

\section{Conclusion}
We presented relativistic calculations of the isotope shifts in Li-like ions. The calculations of the mass shifts were performed by merging the perturbative approach with the CI-DFS method. These calculations confirm our previous results obtained by the CI-DFS method \cite{Kozhedub_2010} and agree with the related MCDF calculations by Li {\it et al.}~\cite{Li_2012} for low- and middle-Z systems. The perturbative method developed in the paper has a potential to be applied for calculations of the QED recoil corrections of the first order in $1/Z$, which define the current theoretical uncertainty.
The CI-DFS calculations of the nuclear size effect performed allowed significant improvements of the theoretical predictions for the field shift constants obtained with the DCB Hamiltonian. The QED corrections to the isotope shifts have been also evaluated. As the result, the most accurate theoretical results for the mass and field shifts of the $2p_{1/2}-2s$ and $2p_{3/2}-2s$ transition energies in Li-like ions have been obtained. For the isotope shifts in elements with $Z$=60, 90, and 92 the nuclear polarization and nuclear deformation effects have been also considered.

\clearpage
\begingroup
\squeezetable

\begin{table*}[H]
\caption{\label{tab1}Contibutions of the individual diagrams to the mass shifts within the Breit approximation in terms of the $K$-factor (in units of 1000 GHz$\cdot$amu) for the $2p_{1/2}-2s$ and $2p_{3/2}-2s$ transitions in Li-like uranium ($Z$=92). }
\begin{ruledtabular}
\begin{tabular}{cccccc|cccc}
&diagram & \multicolumn{4}{c|}{$2p_{1/2}-2s$} & \multicolumn{4}{c}{$2p_{3/2}-2s$}\\
\hline
&&$K_{\rm{NMS}}$ &$K_{\rm{SMS}}$ & $K_{\rm{RNMS}}$ &$K_{\rm{RSMS}}$  &$K_{\rm{NMS}}$ &$K_{\rm{SMS}}$ & $K_{\rm{RNMS}}$ &$K_{\rm{RSMS}}$\\[2mm]
\hline
Zeroth order in $1/Z$ && -3629.93 & -4925.25 &3930.03 & 3929.40&-6989.42 & -2656.60& 6763.01 & 793.10    \\[2mm]
First order in $1/Z$ & a  & 20.78 &285.46 & -72.75 & -254.97 &287.97 & 150.77 & -298.83 & -46.61 \\[2mm]
& b & -3.32 & 13.69 &-0.84 & -12.53  &24.32  &-4.17 &-15.31 & 1.30 \\[2mm]
& c & 0 &2.46 &0  &-15.03 & 0& -14.38 & 0& -1.01\\[2mm]
& d & 0 & -0.86 & 0     & 4.93 &0 & 4.77&0 & 0.35\\[2mm]
& e & -93.15  & 0  &70.29 &0 & -80.11 & 0 &60.55 & 0\\[2mm]
& f & 30.27  &0& -25.07 & 0 &93.16 & 0 &-72.27 & 0 \\[2mm]
& g &0 & 23.68 &0 & -14.82 & 0& 13.59 & 0 & -9.18\\[2mm]
& h   &0& -24.94 & 0& 25.15 & 0& -40.23 & 0 & 26.46\\[2mm]
Sum  & &-3675.35&-4625.76 &3901.66 & 3662.13 & -6664.08&-2546.25 &6437.15&764.41 \\[2mm]
\end{tabular}
\end{ruledtabular}
\end{table*}

\begin{table*}[H] 
\caption{\label{tab2}Mass shift contributions in
terms of the $K$-factor  
(in units of 1000 GHz$\cdot$amu) for the $2p_{1/2}-2s$ transition in Li-like ions.} 
\begin{ruledtabular}
\begin{tabular}{cccccccc}
Contribution &  &&& &&\\
 & {Si$^{11+}$} & {Ar$^{15+}$} & {Zn$^{27+}$} &{Nd$^{57+}$} 
& {Hg$^{77+}$} &{Th$^{87+}$} &{U$^{89+}$} \\            
\hline \\[1mm]
{\rm NMS}\\
Zeroth order in $1/Z$  & -0.628  & -1.742 & -14.370 &-322.61 & -1514.95
& -3141.8   &-3629.9 \\
First order in $1/Z$   & -3.628 & -4.710  & -8.188 & -19.29 &
-29.99&-41.3 &-45.4 \\
Second
and higher orders in $1/Z$ & 0.497& 0.529&0.696& 1.99&5.14&   9.3  &10.5 \\
\hline

{\rm RNMS} & \\
Zeroth order in $1/Z$ &0.630 &  1.747 &
14.455 & 330.22 & 1593.95 & 3382.6 & 3930.0\\
First order in $1/Z$  &-0.054 &  -0.117
&-0.565 & -5.76 & -17.83 & -27.0&-28.4\\
Second and higher orders in $1/Z$ &  -0.012&   -0.021&-0.071& -0.67&  
-2.79&-6.1    & -7.2\\
\hline

{\rm NMS plus RNMS} & \\
This work &-3.195&-4.314&-8.043&-16.12&33.53&175.7&229.7\\
Kozhedub {\it et al.} \cite{Kozhedub_2010} &-&-&-8.054&-16.42&-&-&227\\
Li {\it et al.} \cite{Li_2012}& -&-& -7.895&-14.49&29.61&-&-  \\
\hline

{\rm SMS} & \\
Zeroth order in $1/Z$ &-55.944 &-93.289 &-269.676 &-1321.48 & 
-3005.89 &-4527.5 &-4925.3  \\ 
First order in $1/Z$  & 14.844  &19.369 &34.518 & 96.22 & 188.64 &276.0 &299.5 \\
                 
Second and higher orders in $1/Z$ &  -0.601& -0.624& -0.763& -1.78
&-3.83&   -6.1&-6.7 \\
\hline

{\rm RSMS} &\\
Zeroth order in $1/Z$  &   1.022     &  2.818 & 22.638& 445.12 & 1807.29
& 3454.7   &3929.4\\
First order in $1/Z$  &  -0.313 & -0.686 &-3.387& -37.38 & -128.63&
  -236.2  &-267.3 \\   
Second and higher orders in $1/Z$ &   0.022& 0.037&0.118& 0.89&  3.00& 
   5.7 &6.5 \\
\hline

{\rm SMS plus RSMS} & \\
This work &-40.970&-72.375&-216.552&-818.41&-1139.42&-1033.4&-963.9\\
Kozhedub {\it et al.} \cite{Kozhedub_2010}&-&-&-216.545&-818.09&-&-&-960\\
Li {\it et al.} \cite{Li_2012} &-&-&-216.7&-819.6&-1119&-&- \\
\hline

{\rm Total MS within the Breit approximation} & \\
This work & -44.165&-76.689&-224.595&-834.53&-1105.89&-857.7&-734.2\\
Kozhedub {\it et al.} \cite{Kozhedub_2010}& -&-&-224.600&-834.51&-1105.6&-&-733  \\
Li {\it et al.} \cite{Li_2012} & -44.16 & -76.69 &  -224.6 & -834.1 & -1090 &-783.2 & - \\
\hline
{\rm{QED}} \cite{Kozhedub_2010} &-0.139 & -0.47 &-5.91 &-213.3 & -1167 &-2583 &-3000 \\    
\hline
{\rm Total MS with QED}\\
This work&-44.304 &-77.16 & -230.51 &-1047.8& -2273 &-3441 &-3734 \\
Kozhedub {\it et al.} \cite{Kozhedub_2010}  &- & -& -230.51 & -1047.8 &-2272 & 
-&-3734 \\

\end{tabular}
\end{ruledtabular}
\end{table*} 

\begin{table*}[H]
\caption{\label{tab3}Mass shift contributions in
terms of the $K$-factor 
(in units of 1000 GHz$\cdot$amu) for the $2p_{3/2}-2s$ transition in
Li-like ions.}
\begin{ruledtabular}
\begin{tabular}{cccccccc}
Contribution & &&& &&\\
 & {Si$^{11+}$} & {Ar$^{15+}$} & {Zn$^{27+}$} &{Nd$^{57+}$}
& {Hg$^{77+}$} & {Th$^{87+}$}  &{U$^{89+}$} \\
\hline \\[1mm]
{\rm NMS}\\
Zeroth order in $1/Z$  &-1.332 &-3.686 &-30.140 &-649.61&-2947.52&
 -6052.0 &-6989.4  \\
First order in $1/Z$   &-3.275 &-3.947 &-4.396 &24.50&133.59 & 
280.7    &325.3        \\
Second
and higher orders in $1/Z$ &0.442&0.434&0.396&-0.12& -2.03&
 -4.8  &-5.7 \\
\hline

{\rm RNMS} & \\
Zeroth order in $1/Z$ &1.101 & 3.055&25.240&571.21&2733.90&
 5811.4   &6763.0\\
First order in $1/Z$  &-0.209 &-0.456&-2.360&-32.05&-135.54&
-280.8 &-325.9 \\
Second and higher orders in $1/Z$ &0.000 & 0.001 &0.013& 0.38 &2.03 &
 4.7&5.5  \\
\hline

{\rm NMS plus RNMS} & \\
This work &-3.273&-4.599&-11.247&-85.69&-215.57&
-240.9&-227.1 \\
Kozhedub {\it et al.} \cite{Kozhedub_2010}&-&-&-11.250&-85.73& -&
-&-228\\
Li {\it et al.} \cite{Li_2012} &-& -& -11.29&-85.77& -218.1&
-&-\\
\hline

{\rm SMS} & \\
Zeroth order in $1/Z$ &-55.363&-91.686&-256.779&-1066.12&-1963.75&
 -2532.9 &-2656.6\\
First order in $1/Z$  & 14.593&18.828&31.871&68.10&95.24& 
 108.1 &110.4 \\
Second and higher orders in $1/Z$ &  -0.570&-0.572&  -0.600&-0.75& -0.80&
-0.7& -0.7 \\
\hline

{\rm RSMS} &\\
Zeroth order in $1/Z$  & 0.361 & 0.990& 7.729&130.47&435.79&
 721.4 &793.1 \\
First order in $1/Z$   &-0.109&-0.232 &-1.078& -8.73&-20.28&
-27.4   &-28.7 \\
Second and higher orders in $1/Z$ &  0.008&  0.013&0.035&  0.12&0.12&
-0.0&-0.1  \\
\hline

{\rm SMS plus RSMS} & \\
This work &-41.080& -72.659& -218.822& -876.91& -1453.68& 
-1731.6&-1782.6\\
Kozhedub {\it et al.} \cite{Kozhedub_2010} &-&-&-218.823& -876.93 & -&
-&-1783\\
Li {\it et al.} \cite{Li_2012} &-&-&-218.8& -877.5& -1434 &
-&-\\
\hline

{\rm Total MS within the Breit approximation} & \\
This work &-44.353& -77.258 &-230.069&-962.60&-1669.25& -1972.5 
&-2009.7                   \\
Kozhedub {\it et al.} \cite{Kozhedub_2010}&-&-&-230.073&-962.66& -1669.5&
 -  &-2010 \\
Li {\it et al.} \cite{Li_2012} &-44.34&-77.27&-230.1&-963.2&-1652&
 -1896&- \\
\hline
{\rm {QED}} \cite{Kozhedub_2010} & -0.133 & -0.46 &-5.60&-195.8 &-1082 & -2444  &-2851 \\
\hline
{\rm Total MS with QED}\\
This work & -44.486 & -77.72 & -235.67 & -1158.4 & -2751 &-4416& -4861\\
Kozhedub {\it et al.} \cite{Kozhedub_2010}  &-&-&-235.68& -1158.4&-2751&-&-4861\\

\end{tabular}
\end{ruledtabular}
\end{table*}

\begin{table*}[H]
\caption{\label{tab4}Mass shift contributions in terms of the $K$-factor (in units of 1000 GHz$\cdot$amu) for the $2p_{1/2}-2s$ and $2p_{3/2}-2s$ transitions in Li-like uranium. The calculations are performed starting with five different potentials.}
\begin{ruledtabular}
\begin{tabular}{cccccccc}
Method & DS $$ &KS & PZ & LDF & Coulomb & CI-DFS \cite{Kozhedub_2010} \\
\hline
{$2p_{1/2}-2s$}\\
Zeroth order in $1/Z$ & -705.68 & -720.12 & -708.90 & -722.25 & -695.80 & \\
First order in $1/Z$ & -29.21 & -13.90 & -25.58 & -11.57 & -41.56 & \\
Second and higher orders in $1/Z$ & -&-&-&-&3.13& \\
Total MS within the Breit approximation & -734.89 & -734.02 & -734.48 & -733.82 & -734.23 & -733\\
\hline
{$2p_{3/2}-2s$}\\
Zeroth order in $1/Z$ &-2023.29& -2018.58 & -2021.51& -2014.99 & -2089.91 & \\
  First order in $1/Z$ & 13.70 &8.75& 11.27& 4.93 & 81.14& \\
  Second and higher orders in $1/Z$ & -&-&-&-&-0.91& \\
Total MS within the Breit approximation & -2009.59 & -2009.83 & -2010.24 & -2010.06 & -2009.68 & -2010\\
\end{tabular}
\end{ruledtabular}
\end{table*}

\begin{table*}[H]
\caption{\label{tab5}Total mass shifts in terms of the $K$-factor
(in units of 1000 GHz $\cdot$amu and in units of eV$\cdot$amu) for the $2p_{1/2}-2s$ and $2p_{3/2}-2s$ transitions in Li-like ions.}
\begin{ruledtabular}
\begin{tabular}{ccccc|ccccc}
Ion& \multicolumn{4}{c|}{$2p_{1/2}-2s$} & \multicolumn{4}{c}{$2p_{3/2}-2s$}\\
\hline
& Total non-QED MS & QED & \multicolumn{2}{c|}{Total MS with QED} &Total non-QED MS & QED & \multicolumn{2}{c}{Total MS with QED}\\
&   & & [{1000 GHz $\cdot$amu}] & [{eV$\cdot$amu}] &&&[{1000 GHz $\cdot$amu}] & [{eV$\cdot$amu}]  \\
{ Be$^{+}$}   &-1.5589  &-0.0003   &-1.5592(3) & -0.006448(1) &-1.5591    &-0.0003    &-1.5594(3) &-0.006449(1) \\
{ C$^{3+}$}   &-5.5812  &-0.0021   &-5.5833(13)& -0.023091(5)   &-5.5838    &-0.0022    &-5.5860(13) &-0.023102(5)  \\
{ O$^{5+}$}   &-11.879  &-0.009    &-11.888(3) &-0.04916(1)   &-11.892    &-0.009     &-11.901(3) &-0.04922(1)  \\
{ Ne$^{7+}$}  &-20.420  &-0.027    &-20.447(5) & -0.08456(2)  &-20.456    &-0.027     &-20.483(5) &-0.08471(2)  \\
{ Si$^{11+}$} &-44.165  &-0.139    &-44.304(20)&-0.18323(8) &-44.353    &-0.133     &-44.486(19) & -0.18398(8) \\
{ Ar$^{15+}$} &-76.69   &-0.47     &-77.16(5)  & -0.3191(2) &-77.26     &-0.46      &-77.72(5) &-0.3214(2) \\
{ Ti$^{19+}$} &-117.79  &-1.27     &-119.06(12)& -0.4924(5) &-119.18    &-1.22      &-120.40(11) &-0.4979(5) \\
{Fe$^{23+}$} &-167.21  &-2.89     &-170.10(22) & -0.7035(9)&-170.12    &-2.77      &-172.89(21) &-0.7150(9) \\
{Zn$^{27+}$} &-224.6   &-5.9      &-230.5(4)  & -0.9533(16) &-230.1     &-5.6       &-235.7(4) & -0.9748(15) \\
{Kr$^{33+}$} &-324.5   &-14.7     &-339.2(8)  & -1.403(3) &-336.9     &-13.8    &-350.7(8) & -1.450(3) \\
{Mo$^{39+}$} &-439.5   &-32.3     &-471.8(15) & -1.951(6) &-463.9   &-30.0      &-493.9(14) &-2.043(6)\\
{Xe$^{51+}$} &-700     &-120      &-820(4)    & -3.391(18)&-778       &-110     &-888(4) &-3.672(17)\\
{Nd$^{57+}$} &-835     &-213      &-1048(7)   & -4.334(29) &-963     &-196     &-1159(7) & -4.793(27) \\
{Yb$^{67+}$} &-1029    &-515      &-1544(15)  & -6.39(6) &-1306    &-473     &-1779(14) & -7.36(6) \\
{Hg$^{77+}$} &-1106    &-1167     &-2273(29)  & -9.40(12) &-1669      &-1082    &-2751(27) & -11.38(11)\\
{Bi$^{80+}$} &-1080    &-1493     &-2573(36)  & -10.64(15) &-1773      &-1395    &-3168(34) & -13.10(14) \\
{Fr$^{84+}$} &-988     &-2054     &-3042(47)  & -12.58(19)&-1899      &-1932    &-3831(44) &-15.84(18)\\
{Th$^{87+}$} &-858     &-2583     &-3441(57)  &-14.23(24) &-1972      &-2444    &-4416(54) &-18.26(22) \\
{Pa$^{88+}$} &-800     &-2785     &-3585(61)  & -14.82(25) &-1992      &-2641    &-4633(58) &-19.16(24) \\
{U$^{89+}$}  &-734     &-3000     &-3734(65)  & -15.44(27) &-2010      &-2851    &-4861(62) &-20.10(26) \\

\end{tabular}
\end{ruledtabular}
\end{table*}

\begin{table*}[H]
\caption{\label{tab6} The non-QED field shift in terms of the $F$-factor (in MHz/$\rm{{fm}^2}$) calculated by formulas (\ref{FS_1}),(\ref{FS_2}), and (\ref{FS_3}) (the methods 1,2, and 3, respectively) 
for the $2p_{1/2}-2s$ and $2p_{3/2}-2s$ transitions
 in Li-like titanium, neodymium, and thorium.}
\begin{ruledtabular}
\begin{tabular}{ccccccc}
&& Ion &\\
& transition &{ Ti$^{19+}$}& {Nd$^{57+}$}& {Th$^{87+}$}\\
\hline
&{ $2p_{1/2}-2s$}&\\
Total theory (method 1)&&  -4.8122$\times 10^4$ & -7.5690 $\times 10^6$ &-1.3698$\times 10^8$\\
Total theory (method 2)&&-4.8122$\times 10^4$ &-7.5690$\times 10^6$ &-1.3698$\times 10^8$ \\
Total theory (method 3)&&-4.8437$\times 10^4$& -7.8988$\times 10^6$ & -1.5022$\times 10^8$\\
Li {\it et al.} \cite{Li_2012}&&-4.844$\times 10^4$&-7.885$\times 10^6$&-1.518$\times 10^8$\\[3mm]
\hline\\
&{$2p_{3/2}-2s$}&\\
Total theory (method 1) &&-4.8251$\times 10^4$ & -7.8313$\times 10^6$ & -1.5186$\times 10^8$ \\
Total theory (method 2)&&-4.8251$\times 10^4$ &-7.8313$\times 10^6$ & -1.5186$\times 10^8$ \\
Total theory (method 3)&&-4.8567$\times 10^4$ & -8.1719$\times 10^6$ & -1.6647$\times 10^8$\\
Li {\it et al.} \cite{Li_2012} &&-4.857$\times 10^4$&-8.157$\times 10^6$&-1.681$\times 10^8$\\[3mm]
\end{tabular}
\end{ruledtabular}
\end{table*}

\begin{table*}[H]
\caption{\label{tab7} Comparison of the non-QED FS contribution to the isotope shift obtained by the direct calculation, $\delta E=E(R_1)-E(R_2)$, and by the calculation using the $F$-factor, $\delta E=F \delta \langle r_{12}^2 \rangle$, for the $2p_{1/2}-2s$ and $2p_{3/2}-2s$ transitions in Li-like neodymium, thorium, and uranium.}
\begin{ruledtabular}
\begin{tabular}{c|c|cc|cc|cc}
& &Direct calculation  && \multicolumn{2}{c}{Calculation using the F-factor} \\
\hline 
&{$\delta \langle r_{12}^2 \rangle$} &  && \multicolumn{2}{c|}{method 1} & \multicolumn{2}{c}{method 3}\\
\hline
                         &        &$2p_{1/2}-2s$ & $2p_{3/2}-2s$ & $2p_{1/2}-2s$ & $2p_{3/2}-2s$ &$2p_{1/2}-2s$ & $2p_{3/2}-2s$\\   
$^{150,142}\rm{Nd}^{57+}$& 1.2709 &-0.9592$\times 10^{7}$ &-0.9923$\times 10^{7}$ &-0.9619$\times 10^{7}$ &-0.9952$\times 10^{7}$ &-1.0039$\times 10^{7}$ & -1.0384$\times 10^{7}$  \\
$^{232,230}\rm{Th}^{88+}$& 0.2056 &-0.2819$\times 10^{8}$ & -0.3124$\times 10^{8}$ & -0.2811$\times 10^{8}$ &-0.3122$\times 10^{8}$ &-0.3090$\times 10^{8}$ & -0.3422$\times 10^{8}$ \\
$^{238,236}\rm{U}^{89+}$ & 0.1638 &-0.2703$\times 10^{8}$ & -0.3019$\times 10^{8}$ & -0.2702$\times 10^{8}$  & -0.3016$\times 10^{8}$ &-0.2975$\times 10^{8}$ & -0.3320$\times 10^{8}$ \\
$^{238,234}\rm{U}^{89+}$ & 0.3272 &-0.5404$\times 10^{8}$ & -0.6034$\times 10^{8}$  & -0.5397$\times 10^{8}$  & -0.6025$\times 10^{8}$ & -0.5942$\times 10^{8}$ & -0.6631$\times 10^{8}$\\
\end{tabular}
\end{ruledtabular}
\end{table*}

\begin{table*}[H]
\caption{\label{tab8}Individual contributions to the field shift in terms of
the  $F$-factor (in MHz/$\rm{{fm}^2}$) for the $2p_{1/2}-2s$ and $2p_{3/2}-2s$ transitions
 in Li-like titanium,
 neodymium, and thorium.}
\begin{ruledtabular}
\begin{tabular}{ccccccc}
&& Ion &\\
{ F-factor contributions} & transition &{ Ti$^{19+}$}& { Nd$^{57+}$}&
{ Th$^{87+}$}\\
\hline
&{$2p_{1/2}-2s$}&\\
Dirac-Fock & & -4.8177$\times 10^4$ & -7.5971$\times 10^6$ & -1.3764$\times 10^8$\\
Breit & & 0.0073$\times 10^4$ & 0.0285$\times 10^6$ & 0.0067$\times 10^8$\\
Electron correlation & &-0.0018$\times 10^4$ & -0.0004$\times 10^6$ &-0.0001$\times 10^8$\\
QED && 1.2597$\times 10^2$  & 3.8630$\times 10^4$ & 7.2961$\times 10^5$\\
\hline
Total theory (without QED) && -4.8122$\times 10^4$ & -7.5690$\times 10^6$ & -1.3698$\times 10^8$\\
Total theory (with QED) &&-4.7996$\times 10^4$  & -7.5304$\times 10^6$ & -1.3625$\times 10^8$\\
\hline
\hline
&{ $2p_{3/2}-2s$}&\\
Dirac-Fock & & -4.8304$\times 10^4$ & -7.8606$\times 10^6$ & -1.5265$\times 10^8$\\
Breit & & 0.0071$\times 10^4$ & 0.0297$\times 10^6$ & 0.0079$\times 10^8$\\
Electron correlation &&-0.0018$\times 10^4$ & -0.0004$\times 10^6$ & 0\\
QED &&  1.2655$\times 10^2$ & 4.2314$\times 10^4$ &  1.0729$\times 10^6$\\
\hline
Total theory (without QED) &&-4.8251$\times 10^4$ &-7.8313$\times 10^6$ &-1.5186$\times 10^8$\\
Total theory (with QED) && -4.8124$\times 10^4$  & -7.7890$\times 10^6$  & -1.5079$\times 10^8$\\
\end{tabular}
\end{ruledtabular}
\end{table*}

\begin{table*}[H]
\caption{\label{tab9}Field shifts in terms of the $F$-factor
(in MHz/$\rm{{fm}^2}$ and in meV/$\rm{{fm}^2}$) for the $2p_{1/2}-2s$ transition in Li-like ions.}
\begin{ruledtabular}
\begin{tabular}{cccccccc}
Ion & ${\langle r^2 \rangle}^{1/2}$ & DF & CI-DFS+Breit & QED & \multicolumn{2}{c}{Total}\\
 &&&&&[MHz/$\rm{{fm}^2}$] & [meV/$\rm{{fm}^2}$] \\
\hline
{Be$^{+}$}   &2.5190 & -1.6767$\times 10^1$       &-1.7064$\times 10^1$ &  0.0009$\times 10^1$    &-1.7055(1)$\times 10^1$ & -7.0534(4)$\times 10^{-5}$ \\
{C$^{3+}$}   &2.4702 & -1.4133$\times 10^2$       &-1.4228$\times 10^2$ &  0.0011$\times 10^2$    &-1.4217(1)$\times 10^2$ & -5.8797(4)
$\times 10^{-4}$ \\
{O$^{5+}$}   &2.6991 & -5.4362$\times 10^2$       &-5.4527$\times 10^2$ &  0.0056$\times 10^2$    &-5.4471(1)$\times 10^2$ & -2.25274(4)$\times 10^{-3}$
\\
{Ne$^{7+}$}  &3.0055 & -1.4840$\times 10^3$       &-1.4862$\times 10^3$ &  0.0019$\times 10^3$    &-1.4843(1)$\times 10^3$ & -6.1386(4)$\times 10^{-3}$\\
{Si$^{11+}$} &3.1224 & -6.5518$\times 10^3$       &-6.5520$\times 10^3$ &  0.0115$\times 10^3$    &-6.5405(3)$\times 10^3$ & -0.027049(1)\\
{Ar$^{15+}$} &3.4028 & -1.9764$\times 10^4$       &-1.9751$\times 10^4$ &  0.0044$\times 10^4$    &-1.9707(2)$\times 10^4$ & -0.08150(1)
\\
{Ti$^{19+}$} &3.5921 & -4.8177$\times 10^4$       &-4.8122$\times 10^4$ &  0.0126$\times 10^4$    &-4.7996(6)$\times 10^4$ & -0.19850(2)\\
{Zn$^{27+}$} &3.9491 & -1.9875$\times 10^5$       &-1.9839$\times 10^5$ &  0.0066$\times 10^5$    &-1.9773(4)$\times 10^5$ & -0.81775(17)\\
{Kr$^{33+}$} &4.1835 & -4.7588$\times 10^5$       &-4.7480$\times 10^5$ &  0.0181$\times 10^5$    &-4.7299(14)$\times 10^5$ & -1.9561(6)\\
{Mo$^{39+}$} &4.3151 & -1.0342$\times 10^6$       &-1.0315$\times 10^6$ &  0.0043$\times 10^6$    &-1.0272(4)$\times 10^6$ & -4.2482(16)\\
{Xe$^{51+}$} &4.7964 & -4.0483$\times 10^6$       &-4.0346$\times 10^6$ &  0.0195$\times 10^6$    &-4.015(3)$\times 10^6$ & -16.605(12)\\
{Nd$^{57+}$} &4.9123 & -7.5971$\times 10^6$       &-7.5690$\times 10^6$ &  0.0386$\times 10^6$    &-7.530(6)$\times 10^6$ & -31.142(25)\\
{Yb$^{67+}$} &5.3215 & -2.0431$\times 10^7$       &-2.0345$\times 10^7$ &  0.0111$\times 10^7$    &-2.023(2)$\times 10^7$ & -83.66(8)\\
{Hg$^{77+}$} &5.4463 & -5.3887$\times 10^7$       &-5.3642$\times 10^7$ &  0.0293$\times 10^7$    &-5.335(8)$\times 10^7$ & -220.6(3)\\
{Bi$^{80+}$} &5.5211 & -7.1652$\times 10^7$       &-7.1319$\times 10^7$ &  0.0388$\times 10^7$    &-7.093(11)$\times 10^7$ & -293.3(4)\\
{Fr$^{84+}$} &5.5915 & -1.0487$\times 10^8$       &-1.0437$\times 10^8$ &  0.0056$\times 10^8$    &-1.038(2)$\times 10^8$ & -429.3(8)\\
{Th$^{87+}$} &5.7848 & -1.3764$\times 10^8$       &-1.3698$\times 10^8$ &  0.0073$\times 10^8$    &-1.362(2)$\times 10^8$ & -563.3(8)\\
{Pa$^{88+}$} &5.8291 & -1.5093$\times 10^8$       &-1.5020$\times 10^8$ &  0.0079 $\times 10^8$   &-1.494(3)$\times 10^8$ & -617.9(12)\\
{U$^{89+}$}  &5.8571 & -1.6574$\times 10^8$       &-1.6494$\times 10^8$ &  0.0087$\times 10^8$    &-1.641(3)$\times 10^8$ & -678.7(12)\\
\end{tabular}
\end{ruledtabular}
\end{table*}

\begin{table*}[H]
\caption{\label{tab10}Field shifts in terms of the $F$-factor
(in MHz/$\rm{{fm}^2}$ and in meV/$\rm{{fm}^2}$) for the $2p_{3/2}-2s$ transition in Li-like ions.}
\begin{ruledtabular}
\begin{tabular}{cccccccc}
Ion & ${\langle r^2 \rangle}^{1/2}$ &DF  &CI-DFS+Breit & QED & \multicolumn{2}{c}{Total}\\
&&&&& [MHz/$\rm{{fm}^2}$] & [meV/$\rm{{fm}^2}$]  \\
\hline
{ Be$^{+}$}    &2.5190  & -1.6765$\times 10^1$&-1.7064$\times 10^1$   &   0.0009$\times 10^1$     &-1.7055(1)$\times 10^1$ &-7.0534(4)$\times 10^{-5}$ \\
{C$^{3+}$}    &2.4702  & -1.4132$\times 10^2$ &-1.4227$\times 10^2$   &  0.0011$\times 10^2$     &-1.4216(1)$\times 10^2$ & -5.8793(4)$\times 10^{-4}$ \\
{O$^{5+}$}    &2.6991  & -5.4359$\times 10^2$ &-5.4527$\times 10^2$   &  0.0056$\times 10^2$     &-5.4471(1)$\times 10^2$ &-2.25274(4)$\times 10^{-3}$\\
{Ne$^{7+}$}   &3.0055  & -1.4841$\times 10^3$ &-1.4864$\times 10^3$   &  0.0019$\times 10^3$     &-1.4845(1)$\times 10^3$ &-6.1394(4)$\times 10^{-3}$\\
{Si$^{11+}$}  &3.1224  & -6.5557$\times 10^3$ &-6.5563$\times 10^3$   &  0.0115$\times 10^3$     &-6.5448(3)$\times 10^3$ & -0.027067(1)\\
{Ar$^{15+}$}  &3.4028  & -1.9793$\times 10^4$ &-1.9781$\times 10^4$   &  0.0044$\times 10^4$     &-1.9737(2)$\times 10^4$ &-0.08162(1)\\ 
{Ti$^{19+}$}  &3.5921  & -4.8304$\times 10^4$ &-4.8251$\times 10^4$   &  0.0126 $\times 10^4$    &-4.8124(5)$\times 10^4$ &-0.19902(2)\\
{Zn$^{27+}$}  &3.9491  & -1.9996$\times 10^5$&-1.9960$\times 10^5$   &   0.0067$\times 10^5$     &-1.9893(4)$\times 10^5$ &-0.82271(17)\\
{Kr$^{33+}$}  &4.1835  & -4.8047$\times 10^5$&-4.7940$\times 10^5$   &   0.0185$\times 10^5$     &-4.7755(14)$\times 10^5$ &-1.9750(6)\\
{Mo$^{39+}$}  &4.3151  & -1.0489$\times 10^6$&-1.0461$\times 10^6$   &   0.0045$\times 10^6$     &-1.0416(4)$\times 10^6$ &-4.3077(16)\\
{Xe$^{51+}$}  &4.7964  & -4.1557$\times 10^6$&-4.1416$\times 10^6$   &   0.0208$\times 10^6$     &-4.121(3)$\times 10^6$ &-17.0431(12)\\
{Nd$^{57+}$}  &4.9123  & -7.8606$\times 10^6$&-7.8313$\times 10^6$   &   0.0423$\times 10^6$     &-7.789(6)$\times 10^6$ &-32.213(25)\\
{Yb$^{67+}$}  &5.3215  & -2.1495$\times 10^7$ &-2.1403$\times 10^7$   &  0.0129$\times 10^7$     &-2.127(2)$\times 10^7$ &-87.97(8)\\ 
{Hg$^{77+}$}  &5.4463  & -5.7968$\times 10^7$ &-5.7694$\times 10^7$   &  0.0374$\times 10^7$     &-5.732(10)$\times 10^7$ &-237.1(4)\\
{Bi$^{80+}$}  &5.5211  & -7.7706$\times 10^7$&-7.7329$\times 10^7$   &   0.0513$\times 10^7$     &-7.682(14)$\times 10^7$ &-317.7(6)\\
{Fr$^{84+}$}  &5.5915  & -1.1512$\times 10^8$ &-1.1454$\times 10^8$   &  0.0078$\times 10^8$     &-1.138(2)$\times 10^8$ &-470.6(8)\\
{Th$^{87+}$}  &5.7848  & -1.5265$\times 10^8$ &-1.5186$\times 10^8$   &  0.0107$\times 10^8$     &-1.508(3)$\times 10^8$ &-623.7(12)\\
{Pa$^{88+}$}  &5.8291  & -1.6800$\times 10^8$&-1.6713$\times 10^8$   &   0.0119$\times 10^8$     &-1.659(3)$\times 10^8$ &-686.1(12)\\
{U$^{89+}$}   &5.8571  & -1.8518$\times 10^8$ &-1.8421$\times 10^8$   &  0.0132$\times 10^8$     &-1.829(4)$\times 10^8$ &-756.4(17)\\
\end{tabular}
\end{ruledtabular}
\end{table*}

\begin{table*}[H]
\caption{\label{tab11} Individual contributions to the isotope shifts for the $2p_{1/2}-2s$ and $2p_{3/2}-2s$ transitions in Li-like $^{150,142}\rm{Nd}^{57+}$ 
(in meV) with $^{150,142}\delta {\langle r^2 \rangle}$=1.36 $\rm{fm^2}$.} 
\begin{ruledtabular}
\begin{tabular}{ccc}
 &  { $\rm{2p_{1/2}-2s}$} & { $\rm{2p_{3/2}-2s}$} \\[1mm]
{ {Main contributions}} \\
Field shift   & -42.57 &  -44.05  \\
Mass shift   & 1.30 & 1.50   \\
\hline
FS plus MS, this work & -41.27 & -42.55 \\  
FS plus MS, Li {\it et al.} \cite{Li_2012} & -41.18 & -42.45 \\[1mm]

{ QED}\\
Field shift  & 0.22 & 0.24\\
Mass shift & 0.33 & 0.30 \\[1mm]

{ Others } \\
Nuclear polarization & 0.32 & 0.33 \\
Nuclear deformation & 0.27 & 0.28 \\[1mm]
\hline
Total IS theory, this work \footnotemark & -40.1(2) &  -41.4(2) \\ [2mm]
Total IS theory, Kozhedub {\it et. al.} \cite{Kozhedub_2010} & -40.1 
& -41.4 \\[2mm]
Total IS experiment, Brandau {\it et. al.} \cite{Brandau_2008} & -40.2(3)(6) &  -42.3(12)(20) \\
\end{tabular}
\footnotetext{The uncertainty of $\delta{\langle r^2 \rangle}$ is not included.}
\end{ruledtabular}
\end{table*}

\begin{table*}[H]
\caption{\label{tab12} Individual contributions to the isotope shifts for the $2p_{1/2}-2s$ and $2p_{3/2}-2s$ transitions in Li-like 
$^{232,230}\rm{Th}^{88+}$, $^{238,236}\rm{U}^{89+}$, $^{238,234}\rm{U}^{89+}$ (in meV) with given values of $\delta {\langle r^2 \rangle}$. The values of $\delta {\langle r^2 \rangle}$ are taken from compilation of nuclear radii (R) in Ref. \cite{Angeli_2013}, $\delta \langle {r_{12}^2 \rangle}=R_1^2-R_2^2$.}
\begin{ruledtabular}
\begin{tabular}{ccc|cc|cc}
&  \multicolumn{2}{c|}{$^{232,230}\rm{Th}^{87+}$} & \multicolumn{2}{c|}{$^{238,236}\rm{U}^{89+}$}& \multicolumn{2}{c}{$^{238,234}\rm{U}^{89+}$}\\[1mm]
&  \multicolumn{2}{c|}{$^{232,230}\delta {\langle r^2 \rangle}$=0.21 $\rm{fm^2}$} & \multicolumn{2}{c|}{$^{238,236}\delta {\langle r^2 \rangle}$=0.16 $\rm{fm^2}$}& \multicolumn{2}{c}{$^{238,234}\delta{\langle r^2 \rangle}$=0.33 $\rm{fm^2}$}\\[1mm] 
 & {\bf $\rm{2p_{1/2}-2s}$} & {\bf $\rm{2p_{3/2}-2s}$} & {\bf $\rm{2p_{1/2}-2s}$} & {\bf $\rm{2p_{3/2}-2s}$} & {\bf $\rm{2p_{1/2}-2s}$} & {\bf $\rm{2p_{3/2}-2s}$}\\[1mm]
\hline
{ {Main contributions}}&&&&\\
Field shift  &   -119.0 & -131.9 & -109.1 &-121.9 &-225.1 & -251.4  \\
Mass shift   & 0.1 & 0.3  & 0.1& 0.3 & 0.2 & 0.6\\
\hline
FS plus MS &-118.9  &-131.6 &-109.0 &-121.6  &-224.9  & -250.8\\  

{ QED}&&&&\\
Field shift  &0.6 &0.9  &0.6 &0.9& 1.2 &1.8 \\
Mass shift & 0.4 & 0.4  & 0.4 & 0.4 & 0.9 & 0.8\\[1mm]

{ Others }&&&& \\
Nuclear polarization &1.6  &1.7 & 1.1  &1.2  & 2.3 &2.6\\
Nuclear deformation & 1.5 &1.5 & -2.2 & -2.4 & -2.4 & -2.7\\[1mm]
\hline
Total IS theory \footnotemark  &-114.8(22) &-127.1(22) &-109.1(31) &-121.5(31)& -222.9(32) & -248.3(33) \\
\end{tabular}
\footnotetext{The uncertainty of  $\delta{\langle r^2 \rangle}$ is not included.}

\end{ruledtabular}
\end{table*}
\endgroup

\begin{acknowledgments}
This work was supported by RFBR (Grants No. 13-02-00630, No. 12-03-01140,  No. 14-
02-31476, and No. 14-02-31316), SPbSU (Grants No. 11.38.269.2014,  No. 11.38.261.2014,  No. 11.42.1225.2014, and  
No. 11.50.1607.2013), DFG (Grant No. VO 1707/1-2), GSI, 
the Helmholtz Association, and SAEC Rosatom.  N.A.Z. acknowledges the financial support 
by the Dynasty 
foundation and by G-RISC. The work of  C.\ Brandau was supported by the 
German Federal Ministry for Education and Research (BMBF) [contract number 06GI7127/05P12R6FAN] and by the Alliance Program of the Helmholtz Association (HA216/EMMI).
\end{acknowledgments}


\end{document}